\documentclass[conference]{IEEEtran}
\IEEEoverridecommandlockouts

\usepackage{cite}

\usepackage{color,soul}

\usepackage{amsmath}
\usepackage{siunitx}
\usepackage{graphicx}
\usepackage{tikz}
\usetikzlibrary{shapes.geometric, arrows, decorations.pathmorphing, hobby}
\usepackage{float}
\usepackage{subcaption}

\def\micron{\unit{\micro m}}
\def\FT#1{
\mathcal{F} \left\{ #1 \right\}
}
\def\FTi#1{
\mathcal{F}^{-1} \left\{ #1 \right\}
}

\begin{document}

\title{Accuracy of propagation-based phase-contrast CT under the projection approximation}



\author{
\IEEEauthorblockN{ Giavanna~Jadick 
and~Patrick~La~Rivi\`ere,~\textit{Member,~IEEE}}
\thanks{ This work was supported by the U.S. National Science Foundation Graduate Research Fellowship under Grant No.\ 2140001.}
\thanks{ G. Jadick (e-mail: giavanna@uchicago.edu) and P. La Rivi\`ere (e-mail: pjlarivi@uchicago.edu) are with the Department of Radiology, University of Chicago, Chicago, IL 60637, USA.
}
}

\maketitle

\begin{abstract}
X-ray phase-contrast imaging has the potential to improve image contrast with lower dose by probing an object's refractive properties as well as its absorptive properties. To reconstruct a phase-contrast image from a raw dataset, a phase retrieval algorithm must be applied to invert the forward model of the image acquisition scheme. Discrete forward modeling presents unique computational challenges due to high x-ray wave field sampling requirements. At the cost of accuracy, approximations are often applied for the sake of simplicity and experimental convenience. One of the most ubiquitous simplifications is the projection approximation, which neglects refractive effects within an object. The approximation's validity decreases when imaging thicker objects or using a detector with higher spatial resolution. For greater accuracy, one might use the multislice model instead.
In this work, we explore the accuracy of the projection approximation in simulated synchrotron micro-CT images. We simulated images using two detector resolutions (0.5 and 2 micron) with either the projection approximation or multislice forward model. Paganin phase retrieval was applied to the resulting datasets, and the final phase-contrast images were reconstructed using filtered back-projection. The 2-micron resolution detector images showed little difference between forward models, indicating the projection approximation is satisfied. With sub-micron resolution, there is a noticeable difference in the phase-contrast images, especially around fine details. An accurate phase retrieval algorithm for such high-resolution data likely requires more detailed forward modeling that avoids making the projection approximation.
\end{abstract}


\IEEEpeerreviewmaketitle  

\section{Introduction}

\IEEEPARstart{X}{-ray} phase-contrast imaging (XPCI) has received increasing interest in recent years due to its potential to improve image contrast with lower dose. This is possible by probing an object’s refractive properties, rather than absorptive properties alone as in traditional x-ray imaging. Such contrast is especially advantageous in biomedical imaging scenarios that benefit from high contrast between soft tissues with similar attenuation coefficients. 

To produce phase-contrast images from a raw detected signal, one must solve the so-called ``phase retrieval problem." A variety of techniques have been developed to encode phase information for this purpose, including edge-illumination, grating-based, speckle-tracking, and propagation-based imaging setups \cite{paganin_chapter_2021}. The latter, propagation-based imaging (PBI), offers the unique advantage of not requiring any special optical equipment \cite{gureyev_refracting_2009}. As one shifts the detector further from the object being imaged, phase differences accumulated by the modulated x-ray wave field are amplified. If the radiation source is sufficiently coherent and the spatial resolution is high enough, as in synchrotron micro-CT imaging, Fresnel fringes begin to show along object edges in the detected intensity. The experimental convenience and single-shot potential of PBI lends itself to tomographic imaging \cite{burvall_phase_2011}. 

However, propagation-based XPCI faces several challenges before clinical realization and optimal implementation. Discrete forward modeling presents unique computational challenges; depending on the energy of the x-ray beam, sub-micron sampling is likely necessary for modeling free-space propagation, even for detectors with relatively low resolution. Further, to solve the phase retrieval problem, one must invert the chosen forward model, which may not have an analytical solution and thus can magnify the computational challenge.
For this reason, several approximations are commonly implemented in XPCI forward models. 

One of the most ubiquitous simplifications is the projection approximation, which neglects refractive effects within an object. This is typically valid when the object's thickness is within the near-field regime, generally at higher energies and lower resolutions \cite{paganin_chapter_2021}. More accurately, one can utilize a ``multislice" forward model, which consists of dividing an object into sufficiently thin slices and accounting for refraction between subsequent layers. Under the projection approximation, the x-ray wave field exiting an object is characterized by the line integrals through the real (phase) and imaginary (absorptive) components of the object's complex index of refraction $n_E(\vec{r}) = 1 - \delta_E(\vec{r}) + i\beta_E(\vec{r})$, where $E$ specifies energy-dependence. If one can recover the complex exit wave from the raw data, under the projection approximation, it is straightforward to use the real and imaginary line integral components for tomographic reconstruction.

In the case that the projection approximation is not valid, it is not immediately obvious how or to what degree the resulting tomographic reconstruction will differ from expectation. With rapidly advancing x-ray imaging technology, achieving improved spatial resolutions for larger fields-of-view, this topic is increasingly relevant \cite{sundberg_1-m_2021, yakovlev_wide-field_2022}. The purpose of this work is to explore the impact of the projection approximation in micro-CT images. 

\section{Methodology}


Propagation-based XPCI micro-CT acquisitions were simulated using the projection approximation and multislice forward models. Simulation parameters were varied to probe conditions where the projection approximation may or may not be valid.
The approximated phase-contrast images were compared to the multislice images to assess how the difference in forward model affected the simulation accuracy. 

\begin{figure}
    \centering
    \newcommand\zin{0}
    \newcommand\zex{4}
    \newcommand\zde{8}
    \newcommand\ZMOD[1]{\zin + #1*\zex - #1*\zin}
    \begin{tikzpicture}[
        scale=0.8,
        axis/.style={very thick, ->},
        plane/.style={thick, -},
        dashline/.style={dashed, -}
    ]
    \draw[axis](-\zin-0.5,0) -- (\zde+0.5, 0) node(zax)[right]{$z$};
    \draw[plane] (\zin, 2) node[above] {$\psi_0$} -- (\zin, -2) ;
    \draw[plane] (\zex, 2) node[above] {$\psi_{z_0}$} -- (\zex, -2) ;
    \draw[plane] (\zde, 2) node[above] {$\psi_{z_0+R}$} -- (\zde, -2) ;
    
    \begin{scope}[decoration={snake,amplitude=.5mm, segment length=3mm, post length=0.5mm}]
        \foreach \y in {-3,..., 2} {
            \draw[decorate,->] (\zin-1, 0.45*\y + 0.2, 0)  --  (\zin-0.2, 0.45*\y + 0.2, 0);
        }
    \end{scope}
    
    \draw [shade, top color=blue, bottom color=white, fill opacity=.2,
        decoration={segment length=1cm,amplitude=1cm}, decorate, rounded corners=0.2cm]
    (\ZMOD{0}, .2) --  (\ZMOD{0.07}, 1) -- (\ZMOD{0.4}, 2) -- (\ZMOD{0.6}, 1.5) -- (\ZMOD{0.75}, 1.6) -- (\ZMOD{0.95}, 1) -- (\ZMOD{1}, 0.4) -- (\ZMOD{0.95}, 0) -- (\ZMOD{1}, -1) -- (\ZMOD{0.8}, -1.5) -- (\ZMOD{0.6}, -1.4) -- (\ZMOD{0.4}, -2) -- (\ZMOD{0.28}, -1.9)  -- (\ZMOD{0.01}, -1) -- (\ZMOD{0}, .2);
    
    \newcommand\zarrloc[1]{\ZMOD{0.9} + \ZMOD{#1*0.3}}
    \newcommand\zarrheight[1]{0.5 + #1*0.5}
    \draw[->, thick, decoration={segment length=6cm,amplitude=2cm},  decorate, rounded corners=1mm]
    (\zarrloc{1.1}, \zarrheight{1}) node[right] {$n_E(x,y,z)$} -- 
    (\zarrloc{0.45}, \zarrheight{0.2}) -- (\zarrloc{0.55}, \zarrheight{0.8}) -- (\zarrloc{0}, \zarrheight{0}) ;

    \foreach \z [count=\i] in {0.25, 0.5, 0.75} {
        \draw[dashline] (\ZMOD{\z}, 2) node[above] {$\psi_\i$} -- (\ZMOD{\z}, -2)  ;
    }
    \draw (\ZMOD{0.5}, -2.1) node[below] {$\psi_{i+1} = h_{z_0/N} * \int_{z_i}^{z_{i+1}} \psi_i  \exp[...] dz $};
    
    \draw (\zde, -2.1) node[below] {$\psi_{z_0+R} = h_{R} * \psi_{z_0}$};
    
    \end{tikzpicture}
    \caption{Schematic of the multislice forward model using $N$ slices, where $\exp[...]$ indicates the line integrals through the object's complex index of refraction as given in Eq. \ref{eq:proj_approx}. }
    \label{fig:multislice}
\end{figure}
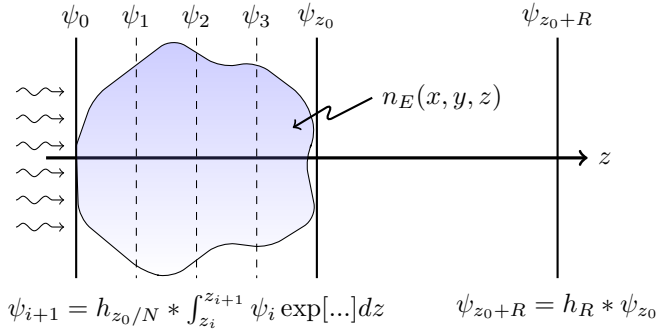

\subsection{Forward models}

We modeled a coherent, monoenergetic plane wave ($n_E \equiv n$, wavenumber $\equiv k$). Under the projection approximation, the wave field exiting an an object can be written in terms of the accrued amplitude ($\beta$) and phase ($\delta$) shifts,
\begin{align}
\label{eq:proj_approx}
    \psi_{z_0} &= \psi_0
    \exp\left[-k \int_0^{z_0} dz ~ \beta(\vec{r}) + i\delta(\vec{r})  \right],
\end{align}
where $z$ is the optical axis and $z_0$ is the location of the object exit plane \cite{paganin_chapter_2021}.
One metric of projection approximation validity is the condition $N_F \gg 1$, with $N_F$ as the Fresnel number
\begin{align}
    \label{eq:NF}
    N_F &= \Delta^2 (\lambda z_0)^{-1},
\end{align}
where $\Delta$ is the spatial resolution, $\lambda$ is the x-ray wavelength, and $z_0$ is the object thickness \cite{paganin_chapter_2021}. 
If this condition is not satisfied, it is more accurate to use the multislice approach (Fig. \ref{fig:multislice}). The object is divided into $N$ slices, each of which should be sufficiently thin to satisfy the condition $N_F \gg 1$. The object exit wave is then computed iteratively.
The wave field at each successive slice $\psi_{i+1}$ is found by multiplying the incident wave $\psi_i$ by the line integrals through the object slice (as in Eq. \ref{eq:proj_approx}) and then applying free-space propagation over the slice width $z_0/N$.

We modeled the free space propagation of a wave as a convolution with the Fresnel operator,
\begin{align}
    \psi_{z_0 + R} &= \psi_{z_0} * ~ h_R = \FTi{ \FT{\psi_{z_0}} \cdot H_R},
\end{align}
where $h_R$ is the spatial-domain and $H_R$ is the Fourier-domain Fresnel operator for propagation distance $R$, 
\begin{align}
    H_R(\nu_x, \nu_y) 
    & = 
    \exp \left[ -\frac{i \pi hc}{E} R (\nu_x^2 + \nu_y^2)\right],
\end{align}
$h$ is Planck's constant, and $c$ is the speed of light in a vacuum. 

After modeling the incident x-ray wave field's modulation through the object, free-space propagation was applied to compute the wave arriving at a detector placed distance $R$ from the object exit plane, $\psi_{z_0+R}$. The final detected signal is equal to the magnitude of this complex wave field.

\subsection{Simulations}


\begin{figure}
\centering
\begin{minipage}[c]{0.45\textwidth}
\centering
    \includegraphics[width=\textwidth]{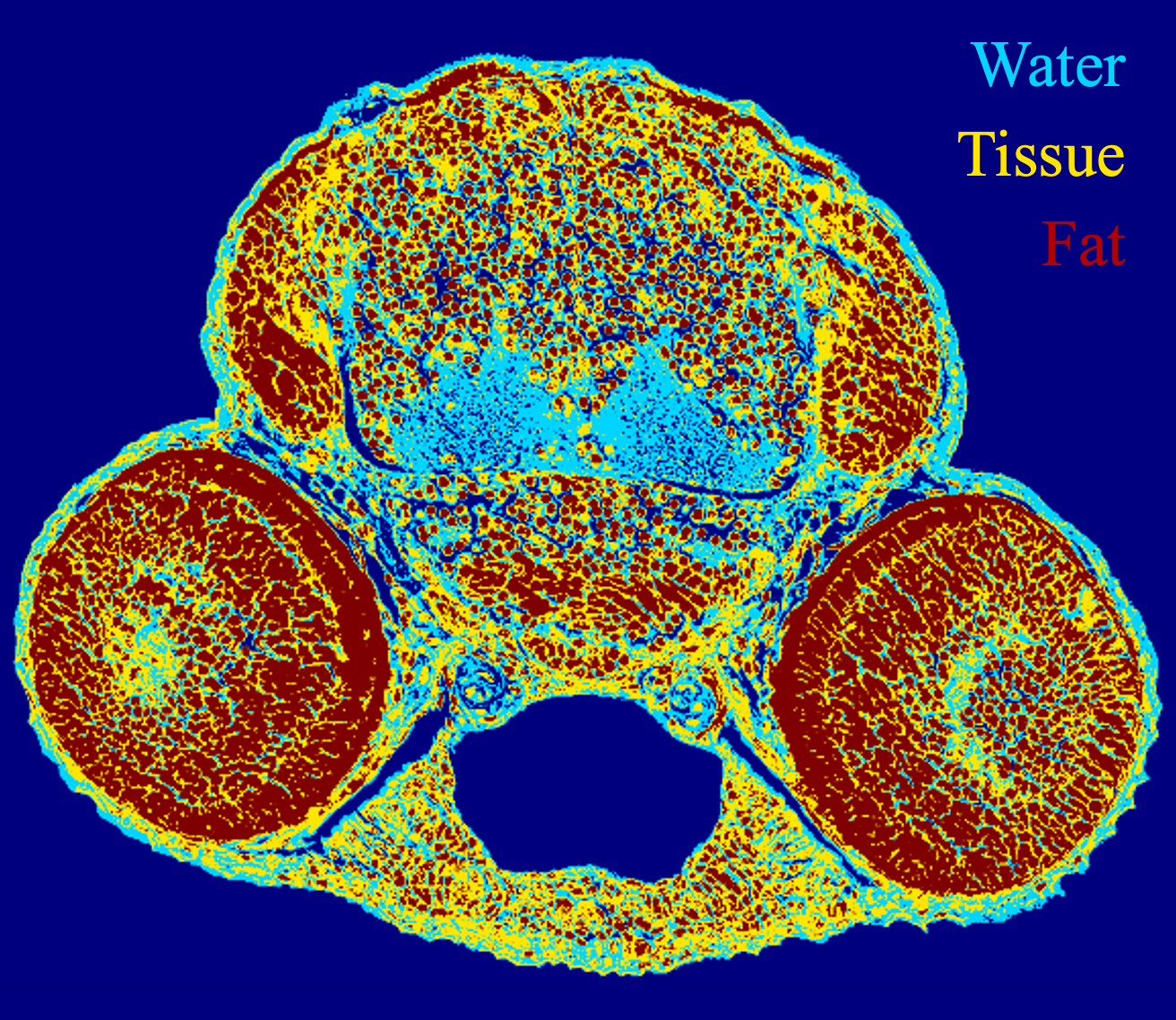}
    \caption{The 5-mm zebrafish phantom used for the micro-CT simulations. }
    \label{fig:phantom}
\end{minipage}
\end{figure}

Parallel-beam, 20-keV micro-CT images of a 5-mm zebrafish phantom (Fig. \ref{fig:phantom}) were simulated using either the projection approximation or multislice modulation with 10 slices. The computational phantom was created from a micro-CT image of a metal-stained zebrafish. The two metals, osmium and uranyl acetate, approximately targeted cell lipids and nuclei, respectively. This facilitated segmentation of the CT image into four materials (air, water, fat, tissue).

Each simulation used a 50-mm propagation distance to the detector and 7-mm beam width. Two detector resolutions were modeled (2.0 and 0.5 \micron), representing tasks that straddle the projection approximation criterion ($N_F=0.8$ and $12.9$, respectively). The wave field was upsampled to a resolution of 0.25 \micron~during each simulation. Projections were acquired at 5000 view angles over a 180-degree rotation. 
Paganin's method for phase retrieval was applied to each projection assuming a single-material object of water \cite{paganin_simultaneous_2002}. This technique utilizes the projection approximation, couples the chosen material's $\delta$ and $\beta$ values, and yields an estimate of the object thickness through each ray. The estimated thickness sinograms were reconstructed into a 5-mm, 4096$\times$4096 array (1.22-\micron~pixel size) using filtered back-projection. 

\section{Results}

\begin{figure*}[t]
\centering
\includegraphics[trim={1mm 6mm 1mm 0}, clip, width=\textwidth]{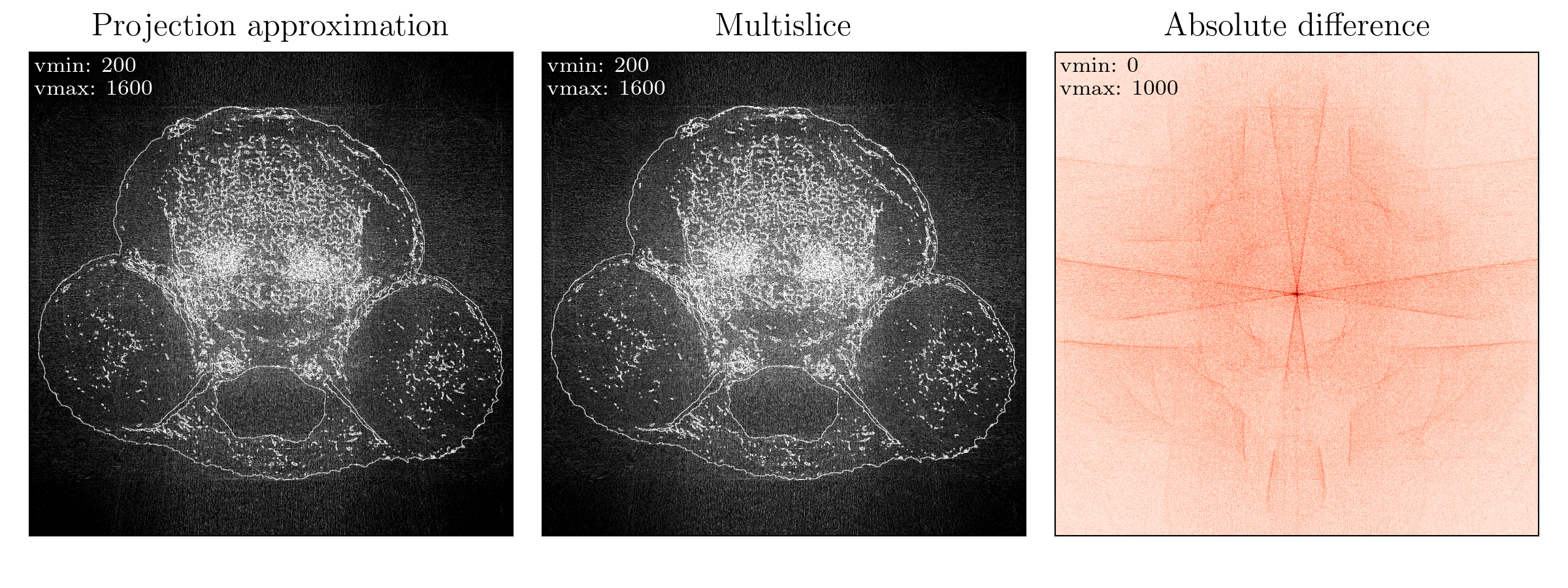}
\caption{For the lower resolution, 2-\micron~detector pixel data, reconstructed images including Paganin phase retrieval and the absolute difference between the images using different forward models.}
\label{fig:recons_paganin_2um}
\end{figure*}

\begin{figure*}[t]
\centering
\includegraphics[trim={1mm 6mm 1mm 0}, width=\textwidth]{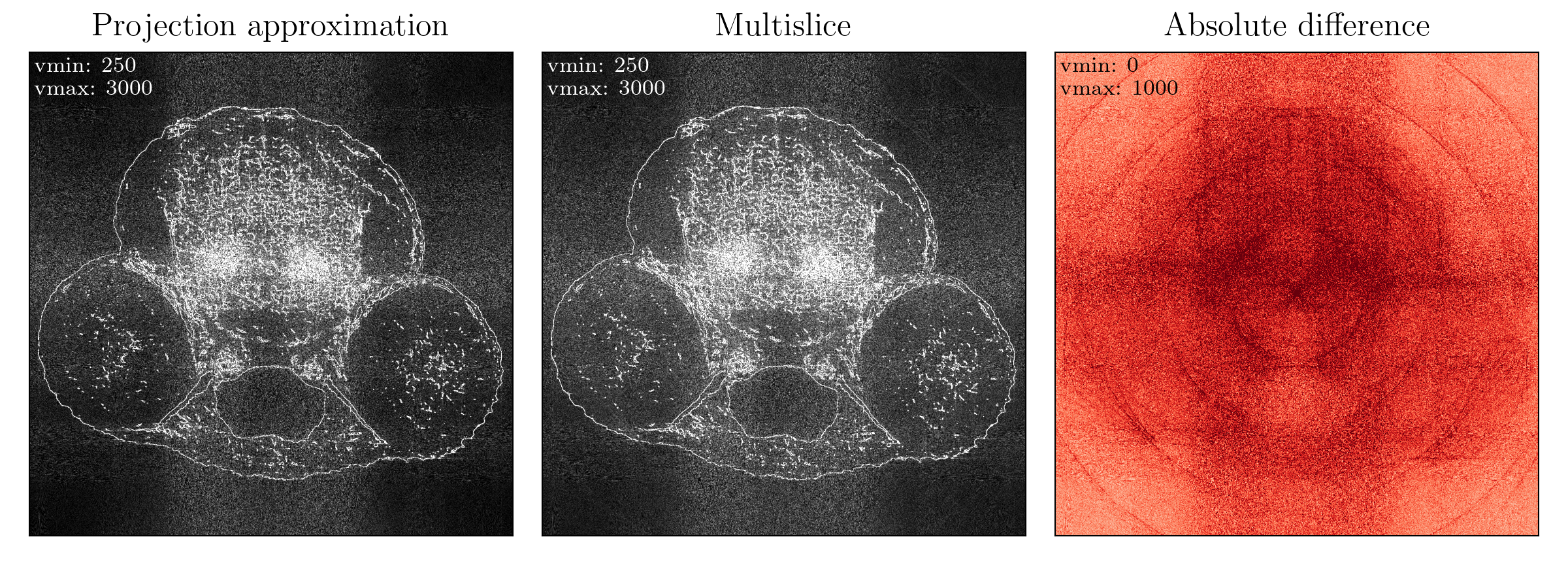}
\caption{For the higher resolution, 0.5-\micron~detector pixel data, reconstructed images including Paganin phase retrieval and the absolute difference between the images using different forward models.}
\label{fig:recons_paganin_highres}
\end{figure*}


\begin{figure}
\centering
\begin{minipage}[c]{0.45\textwidth}
\centering
    \includegraphics[trim={3mm 3mm 0 3mm}, width=\textwidth]{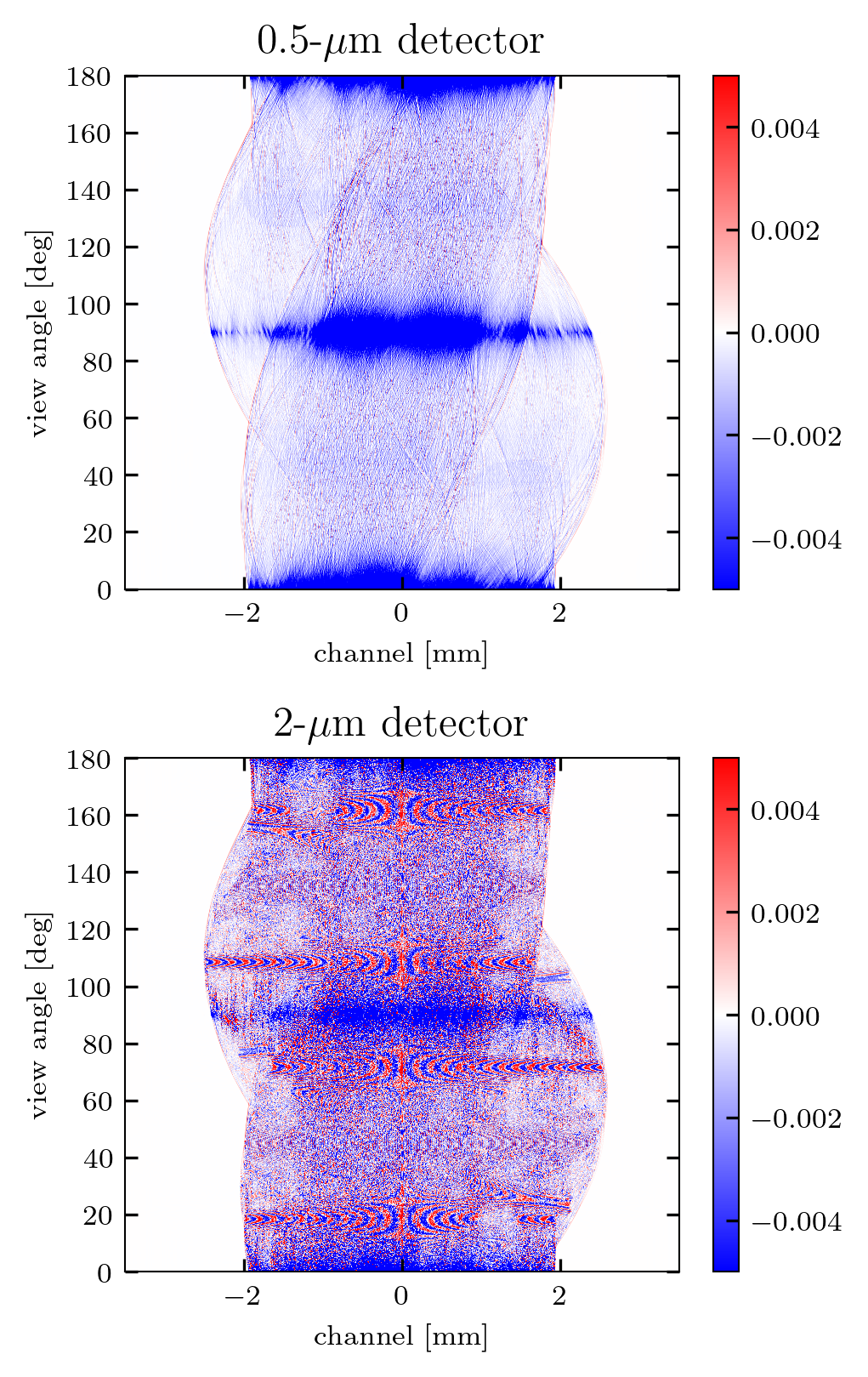}
    \caption{
        Difference in exit wave intensity (projection approximation -- multislice) for the two resolutions. 
    }
    \label{fig:sinodiff_0mm}
\end{minipage}
\end{figure}

Figures \ref{fig:recons_paganin_2um} and \ref{fig:recons_paganin_highres} show the reconstructed phase-contrast images simulated with and without the projection approximation at the two detector resolutions. The images using the 2-\micron~detector resolution show little difference whether or not applying the projection approximation. This indicates that the conditions for the projection approximation are likely satisfied, as predicted by $N_F = 12.9 \gg 1$. 

The 0.5-\micron~ images show a greater difference between the two forward models, especially in regions of the phantom with fine details. This is expected, as the smaller Fresnel number ($N_F = 0.8$) for this detector indicates that refractive effects within the 5-mm phantom thickness are not insignificant, and propagation-based phase effects manifest as Fresnel fringes along the edges of structures.
This increased intensity at structure edges appears to cause bright streaks in the reconstructed images along lines passing through many fine structures. This is slightly more noticeable in the multislice image but apparent with the projection approximation as well.

In all reconstructions, intricate fatty regions (membranes) display good contrast, but it is challenging to resolve tissue regions (cellular nuclei) on the water background. There is a hazy increase in intensity towards the center of the images, particular with the multislice simulations. This may be due to the images not satisfying the conditions for Paganin phase retrieval. In addition to the projection approximation, Paganin's method assumes the object is made of a single material object and the detector is in the near-field regime. 

Figure \ref{fig:sinodiff_0mm} shows the exit wave intensity difference between the two forward models. The 0.5-\micron~sampling is sufficiently high to resolve striations in the sinogram that appear along the edges of fine structures, while the 2-\micron~sampling shows aliasing artifacts due to block averaging over the high-frequency fringes around these fine structures.

\section{Discussion}

The projection approximation is widely implemented in propagation-based XPCI forward models due to its simplicity, convenience for phase retrieval, and applicability to tomographic reconstruction. The approximation is valid when the refractive effects within an object are negligible at the target resolution. As modern imaging tasks utilize larger objects and smaller detector pixels, accurate modeling may require a more complex approach such as the multislice forward model. In this work, we assess how different forward models affect simulated micro-CT images. We find that with a sub-micron resolution detector, images simulated with the projection approximation noticeably differ from those using the multislice approach.
The simulated image pairs offer qualitative insight into the consequences of violating this forward model assumption.

We applied Paganin phase retrieval to the simulated CT acquisitions, as it is widely utilized in experimental settings due to its convenience of only requiring a single input image \cite{paganin_simultaneous_2002}. More generally, since a phase retrieval algorithm aims to recover two unknowns (phase and amplitude components of an object), two independent measurements are required. These are typically acquired at different propagation distances, which is experimentally inconvenient and error-prone \cite{burvall_phase_2011}. To perform phase retrieval from just one image, Paganin's algorithm applies simplifications that reduce its quantitative accuracy, including the projection approximation. Our results match the expectation that when the projection approximation is not satisfied (0.5-\micron~detector), there is a discrepancy in Paganin phase-contrast images simulated with the multislice forward model. Paganin's algorithm also assumes that an object comprises a single material and that the propagation distance to the detector is in the near-field regime. While the zebrafish phantom has a composition similar to that of water, it also included fat and soft tissue. Although the 2.0-\micron~detector images show only small differences between the projection approximation and multislice forward models, there are imperfections in the phase retrieval that are apparent in both images regardless of resolution. This may be due to not satisfying the additional assumptions of Paganin's algorithm.

Tomographic reconstruction algorithms such as filtered back projection require a set of line integrals through an object as their input. For a phase retrieval algorithm that rightfully applies the projection approximation, one expects to recover the line integrals through an object's $\delta$ and $\beta$ components. Since Paganin's phase retrieval assumes a single material, it recovers line integrals through the object thickness. If the assumptions for the phase retrieval algorithm are not satisfied, propagation-based artifacts such as residual Fresnel fringes will contaminate the recovered line integrals. We see this effect especially in the 0.5-\micron~detector images, where the assumptions of Paganin's algorithm are less valid. Consequently, the reconstructed phase-contrast image appears to have increased intensity along lines passing through highly detailed structures.


\section{Conclusion}

For accurate propagation-based XPCI forward modeling, the multislice approach may be used for high-resolution imaging of relatively thick objects. To reconstruct phase-contrast images for such datasets, phase retrieval methods should account for the refractive effects occurring within the object and minimize their use of other simplifying assumptions. At lower resolutions, the projection approximation is often sufficient.

\section*{Acknowledgment}

The authors would like to thank Phillip Vargas for creating the zebrafish phantom used in the micro-CT simulations.


\bibliography{references.bib}{}
\bibliographystyle{ieeetr}


\end{document}